# Light-triggered on/off-switchable bioelectronic FET device

Vikram Bakaraju[1], E. Senthil Prasad [2] and Harsh Chaturvedi[3,a]

[1]Department of Physics, Indian Institute of Science Education & Research (IISER), Pune, MH, 411021, India

[2]Institute of MicrobioTechnology (IMTECH), Chandigarh, India

[3]Indian Institute of Technology (IIT), Guwahati, Assam, India

a) harshc@iitg.ac.in

**Abstract**:

We report fabrication of an optically active, bio-electronic device based on thin film of purple membrane and single walled carbon nanotubes (SWNT). Two dimensional (2D) crystals of photoactive bacteriorhodopsin forms the optical center of the purple membrane where as pure SWNTs provides the necessary electronic support to the complex. Electro-optically functional and stable, hybrid complex was prepared using surface functionalization of SWNTs with indigenous, batch-process synthesized purple membrane. Raman spectra of the hybrid complex shows significant charge transfer and surface functionalization of SWNTs. Optically active, field effect transistor based on the prepared thin film of bio-nano hybrid complex is fabricated using direct laser lithography and conventional lift off technology. Significant optical doping is observed in the fabricated field effect transistor. Fabricated devices show repeatable stable performance with well-controlled optical and electronic gating. Device shows essentially n-type FET characteristics and transistor is ON for positive gate voltages. However, same n-type FET shows complimentary p-type characteristics under visible light illumination, with transistor being ON for negative gate voltages. Significant optical doping, photo-conductivity and optical switching were observed.

**Introduction**:

Research on Bio-Nano hybrid materials is being actively pursued for potential application in developing sustainable, novel electro-optical devices and biological sensors.(Bräuchle, Hampp et al. 1991, Hampp 2000, Huang, Wu et al. 2004) Donor acceptor system based on SWNTs functionalized with inorganic, organic polymers,(Borghetti, Derycke et al. 2006) biological DNA,(Shim, Shi Kam et al. 2002, Keren, Berman et al. 2003, Staii, Johnson et al. 2005) protein etc have been widely reported in pursuit of developing novel devices, actuators, sensors based on these hybrid materials.(Singh, Pantarotto et al. 2005) Devices such as field effect transistors, sensors, rectifiers based on SWNTs functionalized with optically active materials like quantum dots, inorganic ruthenium dyes or optically active molecules has been reported.(Katz and Willner 2004) Photoactive proteins/molecules bind covalently or non-covalently with SWNTs, to form stable donor-acceptor system.(Guldi, Rahman et al. 2005, Sharma, Prasad et al. 2015) Non-covalent functionalization of SWNTs is generally preferred for electro-optical devices, as it preserves the electronic nature of pristine SWNTs in the functionalized hybrid complex. Bioelectronics explores use of electronic/optically active, functional biological molecules (chromophores, proteins, etc.) as an active material for electronic or photonic, plasmonic devices or sensors. Diverse devices are being proposed, based on SWNTs functionalized with optically active, biological molecules.(Barone, Baik et al. 2005, Wang 2005) Bacteriorhodopsin is a rare optically active protein, widely proposed for its technological application in developing various electronic and photonic devices such as optical data storage, electro--optic memory, logic gates, photo-chromatic and holographic systems. Recent reports of functionalization of Bacteriorhodopsin with nanoparticles (quantum dots, nanotubes) shows active interest in using functionalized bio-nano hybrid complex for developing novel photovoltaic, electro-optic devices.(Jin, Honig et al. 2008, Lu, Wang et al. 2015) Here in, we report fabrication of optically active, field effect transistor (FET) based on SWNTs functionalized with bacteriorhodopsin. Fabricated FET device shows interesting electro-optical properties such as intrinsic optical doping, photoconductivity and optical gating.

Bacteriorhodopsin forms the optical centre of the Purple membrane. Purple membrane is made up of 75% protein and 25% lipid.(Lozier, Bogomolni et al. 1975) Only protein present in the purple membrane is bacteriorhodopsin, which acts as a light driven proton pump.(Lozier, Bogomolni et al. 1975) Photocycle of bacteriorhodopsin in the PM has been well characterized. PM remains stable up to 80 C, has buoyant density of 1.18 g/cm$^3$, Refractive index of 1.45 – 1.55 and its natural stable crystalline structure makes purple membrane excellent two dimensional optical material for the development of bio-electronics.(Xu, Bhattacharya et al. 2004, Wang, Knopf et al. 2006) Optoelectronic and photonic applications demand high quantity (several milligrams) thin films of purple membrane. For large-scale incubation of bacteriorhodopsin, Halobacterium cultivation was performed in a 7 liter photo-bioreactor. Purified purple membrane were characterized using UV-Vis spectrophotometer by taking ratio of absorptions at 280 nm and 560 nm. Along with high yield of 14.4 mg/l, extracted PM shows high quality estimated using absorption peaks, to be around ∼ 2.0 to 2.1. (**Figure 1**) 2 mg/ml of this extracted, protein aliquot was used for further functionalization with SWNTs.

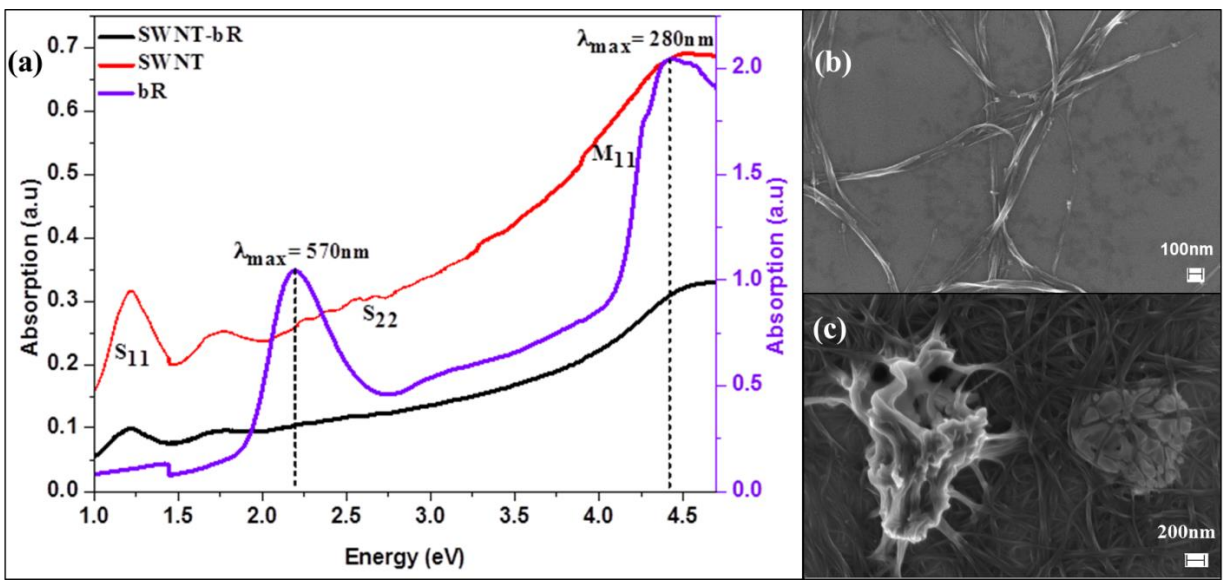

Figure 1 (a) Absorption spectra of pure SWNT (red) and bacteriorhodopsin (bR) (blue) control solutions and that of functionalized SWNT –bR complex. (b) SEM image of pure SWNT (c) functionalized with bR.

Proposed Bio-nano devices based on SWNTs are limited due to inherent hydrophobic nature of SWNTs. However, aqueous solutions of SWNTs have been reported using surfactants.(Vaisman, Wagner et al. 2006) We have used biocompatible phospholipid polyethylene glycol amine as surfactant and established protocol to prepare aqueous solution of pure SWNTs. (Liu, Tabakman et al. 2009)

**Materials and Methods**:

Aqueous solutions of SWNTs were prepared using well-established protocol (Sharma, Prasad et al. 2015). Ultrapure (99.5%) SWCNTs, purchased from NanoIntegris in the form of sheets (batch number: P10-126) were dispersed in water using 1,2-Distearoyl - *sn* - Glycero - 3 - Phosphoethanolamine -*N*- Amino (polyethylene Glycol) 2000] (Ammonium Salt) (Product code: 880128P), bought from Sigma Aldrich, as the surfactant. Stable aqueous solution was obtained for the ratio of 1:7.5 (1 part of SWCNT with 7.5 parts of PlPEG amine). Measured absorption spectra (UV-vis NIR) of the prepared aqueous solutions, do not show any discernible changes in the concentration of SWNTs, over weeks. Aqueous solution of $10^{-2}$M concentration of synthesized Purple membrane and dispersed SWNT solution was prepared and further used for fabricating devices.  Light based separation of pure and functionalized SWNTs has recently been (Gopannagari and Chaturvedi 2015)(Chaturvedi and Poler 2007).  We have also recently communicated technique of optical separation of SWNTs functionalized with the purple membrane.(Sharma, Prasad et al. 2015) Dispersed solution of SWNTs and PM under broadband illumination shows rapid rate of aggregation.(Sharma, Prasad et al. 2015) Stable aqueous solution of SWNTs were illuminated by broadband mercury lamp for 3 hrs. Aggregates flocs of SWNTs were separated from the dispersed supernatant using centrifugation. Aggregates of functionalized SWNT were then drop casted onto the fabricated device and further characterized using Raman spectroscopy. Figure 1b shows SEM images of pristine (top) as well as that of separated aggregates of SWNT functionalized with bacteriorhodopsin. Bacteriorhodopsin crystal in purple membrane is expected to be ~ 500 nm; where as pristine SWNTs are ~ 1-2 micron in length with diameters varying from 0.7 nm upto 1.8 nm. HR-SEM  images have been taken using ZIESS system with back scattering configuration. Images distinctly show functionalization with SWNTs extended out from the matrices of 2- dimensional bacteriorhodopsin-protein complex.

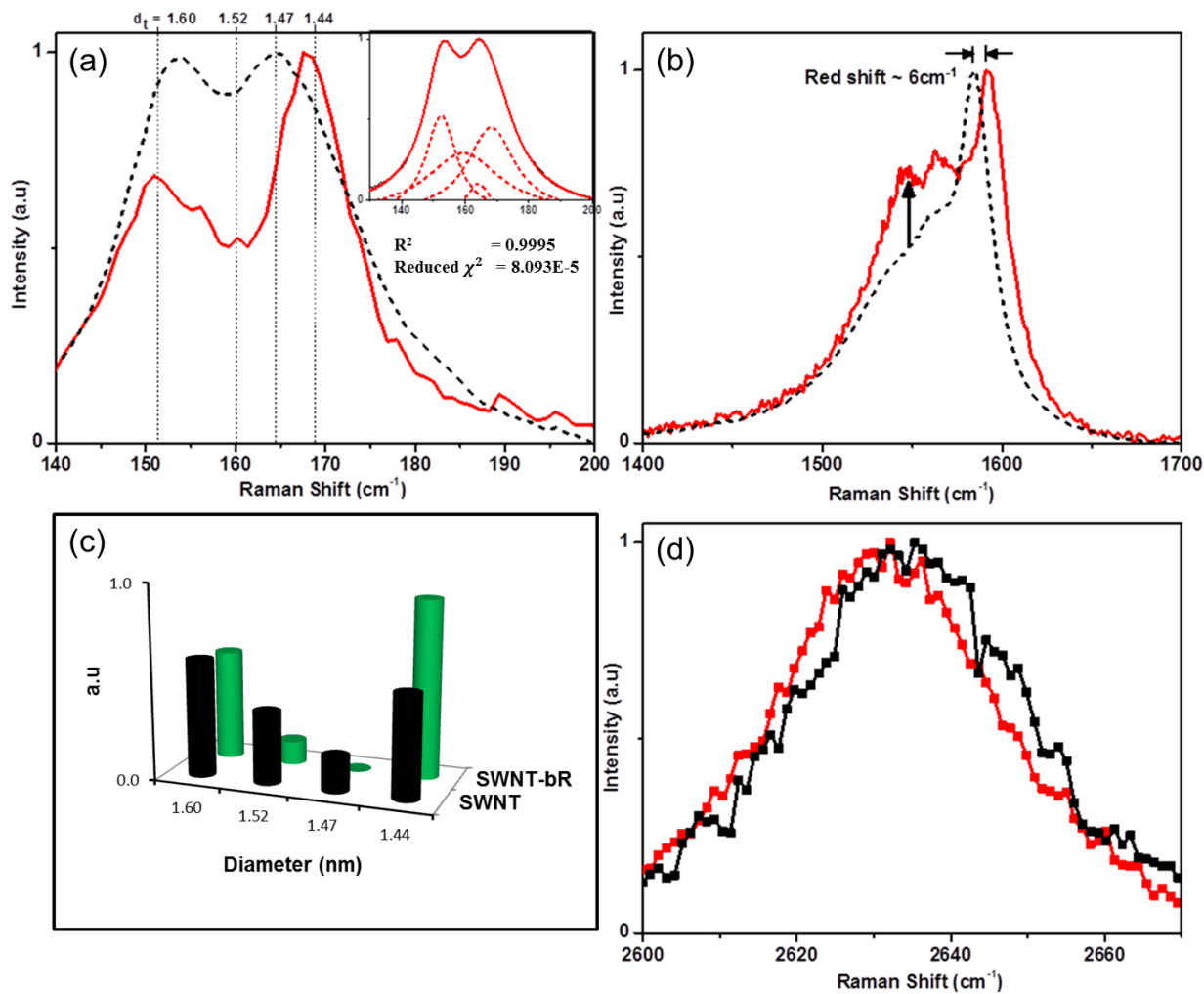

Figure 2: Raman spectra showing (a) RBM of SWNT (black) and functionalized complex of SWNT-bR(red) showing stable preferential binding with specific diameters. (b) G band showing changes in $G^+$ (1591 cm$^{-1}$) and $G^-$ (1540cm$^{-1}$) peak of SWNT-PM Complex as compared to pure SWNTs (c) Histogram shows preferential binding , enrichment SWNTs of specific diameters (d) G' band shows slight red shift (3cm$^{-1}$).

Hybrid complex of SWNT functionalized with bacteriorhodopsin was further characterized using Raman spectrometer with 632 Red laser line. Raman spectra (**Figure 2**) of the functionalized samples show stable and preferential binding of PM with SWNTs of specific diameters. RBM of Raman spectra essentially depends on the diameter of the SWNTs which may be calculated using the relation $\omega_{RBM} = (\alpha_{RBM}/d) + \alpha_{bundle}$. Where, $\alpha_{RBM}, \alpha_{bundle}$ are constants and $d$ is the diameter of the SWNT corresponding to the RBM peak frequency ($\omega_{RBM}$).(Bachilo, Strano et al. 2002)

Histogram (Figure 2 c) as calculated from RBM of Raman spectra (Figure 2 a) shows enrichment of specific diameters of SWNTs in the hybrid complex. Discernible changes in G band of functionalized SWNT is observed as compared to pristine SWNTs. Changes in $G^-$ band are related to metallicity of the SWNTs and have been shown to vary with the changes in the electronic properties of the nanotubes. Functionalized SWNTs exhibit either Red/Blue shift depending on the charge (n type/ p type) donated from the functional molecules onto SWNTs.(Rao, Eklund et al. 1997, Fantini, Jorio et al. 2004) Red shift of ~ 6 $cm^{-1}$ is observed in SWNTs functionalized with PM. This red shift can correspondingly relate to charges being donated on to the nanotubes.(Rao and Voggu 2010) Using the similar approach as cited reference, red shift in our devices corresponds to n ≈ $3*10^6$ $cm^{-2}$. Effect of this charge donation was further evident as "optical doping" of the fabricated device.

Devices were fabricated using standard lift off technology and direct laser lithography to pattern metallic pads on p-type doped silicon wafer (purchased from Semiconductor Wafer. Inc). Metallic patterns acts as source and drain whereas sample holder (chuck) acts as back gate of the field effect transistor. Optically separated, thin film of purple membrane and SWNTs acts as the functional conducting channel.

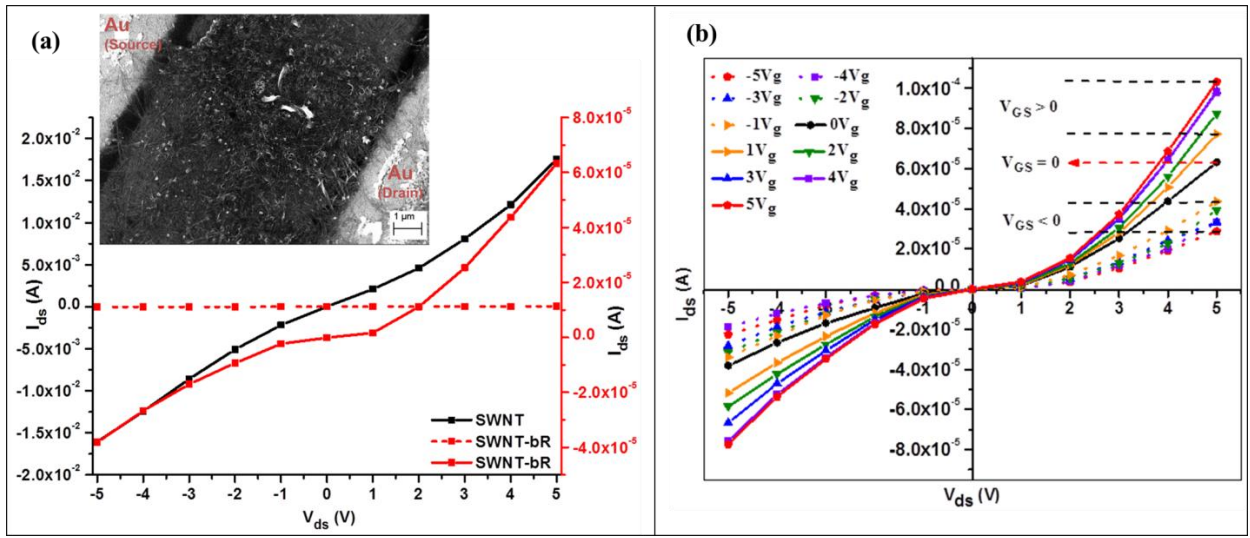

Figure 3: (a) Comparative I-V characteristics of FET based on pure SWNTs and SWNTs functionalized with bacteriorhodopsin (bR). double y axis indicates considerable change in the $I_{ds}$ before and after functionalization. SEM image of device fabricated is shown as an inset. (b) I-V characteristics of the fabricated FET based on functionalized SWNT-bR complex for a range of gate bias applied [-5$V_{gs}$, +5$V_{gs}$].

Figure 3 (inset) shows the HR SEM image of the fabricated FET based on SWNT functionalized with PM. Devices were electro-optically characterized using integrated probe station with micromanipulator attached with Keithley 4200 SCS Semiconductor characterization system. Functionalized SWNTs shows significantly lower ($10^{-2}$) current than pristine SWNTs. This significant decrease in current is most likely expected due to binding with non-conducting purple membrane. Although, pure bacteriorhodopsin as control do not show any significant conductivity as is well-reported; pure SWNTs, as produced includes both metallic and semiconducting SWNTs, hence showing higher conductivity, linear resistivity and not much of gate control. Whereas, Figure 1(a) shows nonlinear resistivity and increase in semiconducting property of the functionalized complex, as compared to pristine SWNTs. Fabricated FET based on PM functionalized SWNT shows much better gate control and switching properties, as shown in Figure 1(b). Devices functionalized with PM show characteristic n-type semiconducting behaviour and significantly enhanced control on current using gate voltages. FET shows higher (1.25 times) current for positive gate voltages as compared to applied negative voltages. Well-controlled decrease in current is observed with corresponding increase in applied negative voltages. Moreover, under light, "optical doping" is observed. Figure 4 compares the electronic characteristics of the devices under dark and light. Control FET based on pristine SWNTs do not show any change in conductivity under illumination by broadband mercury lamp. However, devices based on SWNTs functionalized with PM under broadband illumination shows p-type characteristics as compared to n-type characteristics under dark. Figure 4 (a) validates discernible changes in the 3-Dimensional I-V characteristics of FET based on SWNT functionalized with PM with respect to both the gate voltage and applied drain and source voltages. When in dark, Device is 'On' for positive gate voltages, however, under light device turns "Off" and vice versa for negative gate voltages.

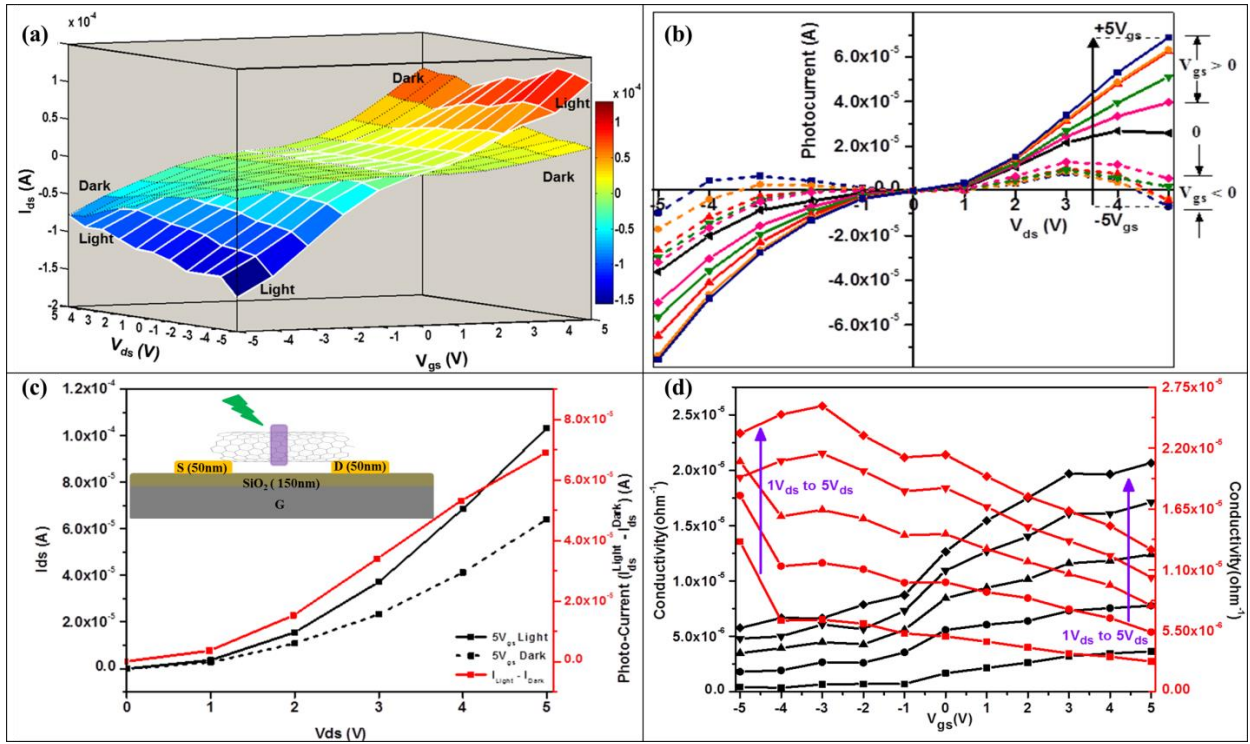

Fig 4: (a) 3-Dimension plot of current-voltage (I-V) response of the device with varying drain – source and gate voltages applied under conditions of dark and light.(b) Plot showing increase in photocurrent with increase in gate voltages. Device do not show any significant change in photocurrent for negative gate voltages. (c) Comparison of current under Light (Solid line)and in Dark (Dashed) at +5 $V_{GS}$ . Photocurrent is shown on secondary Y-Axis. (c) Plot showing "Optical Doping" with device switching from n-type (black) under dark to p-type under light (red).

Changes in photocurrent ($I_{Light} - I_{Dark}$) with variation in drain-source voltages are plotted in Figure 4(b). Device do not show any appreciable photocurrent for negative gate voltages, however device turns 'On' for positive gate voltages. With increase in positive gate voltages ($V_{GS}$), corresponding increase in photocurrent is observed. Difference in drain and source current under light and in dark, at constant voltage of +5V is shown in Figure 4(c). Photocurrent consistently increases with corresponding increase in applied voltages. Measured change in photocurrent is shown on the secondary Y axis.

Both under light and in dark, device shows characteristic FET switching with distinct active and saturation regions. With change in gate voltages device distinctly turns On and Off. Figure 4(d) shows both "Optical doping" and "Electronic gating" of the device. In dark, device turns "On" for positive gate voltages and current increases with increase in positive gate voltages. Whereas under light the device is "Off" for positive gate voltages and turns "On" for negative gate voltages.  Table 1 illustrates the electronic characteristics of the device under dark and under illumination by broadband frequency mercury lamp.

| **Electrical Parameter (SWNT-bR)** | **Dark** | **Light** |
|---|---|---|
| **Threshold Voltage ($V_t$)** | +1.75 V | +1.8 V |
| **Mobility (μ)** | ~433.92 $cm^{-2} V^{-1} S^{-1}$ | ~471.15 $cm^{-2} V^{-1} S^{-1}$ |
| **Photo Current ($I_{Light} - I_{Dark}$)** | - | ~ 68.9 μA |
| **Conductivity (G)** | ~ 2.0 * $10^{-5}$ (S/m) (+5$V_{gs}$) | ~ 2.30 * $10^{-5}$ (S/m) (+5$V_{gs}$) |
| **Majority Carriers** | n-type | p-type |

Table 1: Comparison of the calculated electronic parametr of the fabricated FET device based on SWNT functionalized with PM in dark and under broadband illumination.

Here in we observe singinificant changes in the electronic properties of the functionalized FET under light as compared to measurements on the same devices in dark. We attribute this to well-established photocycle and proton transfer properties of PM. Under optical illumination, signicant conformal as well as electronic changes in PM is well reported. Under light 2 D crystal of bacteriorhodopsin acts as a proton pump, we believe there by converting n-type functionalized SWNT FET into p-type devices. SWNTT FET devices works on the principle of schottky barriers at the metal semiconductor contact. SWNTs are generally considered to be ambiopolar, with symmetrical changes in current for both positive and negative gate voltages. However, since electronic transport through the device significantly depends on the Schottky barrier; any modulation in electronic concentration is expected to cause significant changes in the electronic property of the device.

In conclusion, we have shown stable, optically functional bioelectronic device based on purple memberane and SWNT complex. Both "optical doping" and significant "optical switching" is observed in fabricated FETs. Devices also show considerable photocurrent and well controlled electronic gating. Well-controlled electro-optical functionality of the device is due to strong interaction and charge transfer between the 2-dimensional optically active bacterirhodopsin and electronic single walled carbon nanotubes, supporting data from Raman spectroscopy. We believe the results discussed here will be important for diverse biophotonic, bioelectronic, biosensing and photo-voltaic applications. Results shown here may also contribute in realizing functional reconfigureable bio-photonic, electro-optical devices with devices showing reversible and controlled change in the type of majority carriers, depending on the condition wether the device is under illumination or is in dark. "Optical doping", "Electro-Optical switching" and "Electronic gating" – the fact that each of these phenomenon reported above are using bio-nano hybrid complex and realized in the same device, promises potential for widespread application in diverse areas from biological sciences to electronic photonic hybrid sensors and devices.

**Acknowledgements:**

Authors are deeply indebted to Ramanujan fellowship (SR/S2/RJN-28/2009) and funding agencies DST (DST/TSG/PT/2012/66), Nanomission (SR/NM/NS-15/2012) for generous grants.